\documentclass{iopart}

\begin{document}

\title[Propagation of Dirac wave functions ...]{Propagation of Dirac wave functions in accelerated frames of reference}
\author[Y-C Huang and W-T Ni]{Yi-Cheng Huang and Wei-Tou Ni}
\address{Department of Physics, National
Tsing-Hua University \par HsinChu, Taiwan 300, R.O.C.}
\begin{abstract}
The first-order gravity effects of Dirac wave functions are found
from the inertial effects in the accelerated frames of reference.
Derivations and discussions about Lense-Thirring effect and the
gyrogravitational ratio for intrinsic spin are presented. We use
coordinate transformations among reference frames to study and
understand the Lense-Thirring effect of a scalar particle. For a
Dirac particle, the wave-function transformation operator from an
inertial frame to a moving accelerated frame is obtained. From
this, the Dirac wave function is solved and its change of
polarization gives the gyrogravitational ratio 1 for the
first-order gravitational effects. The eikonal approach to this
problem is presented in the end for ready extension to
investigations involving curvature terms.

\end{abstract}
\section{Introduction}
According to the Einstein's Equivalence Principle, inertial
effects gives the first-order gravity effects. As a result, many
efforts have been made in studying the inertial effects and
relating them to gravity \cite{a,b,c,o,d}. In previous papers of
our group, we have studied the equations of motion and
trajectories of scalar particles in an accelerated frame and in a
gravitational field \cite{b,c,p}. In 1990, we obtained the Dirac
equation in an accelerated frame and studied its various inertial
effects \cite{d}. In this paper, the propagation of Dirac wave
functions in accelerated frames with an emphasis on spin effects
will be studied, and the discussion of the gyrogravitational ratio
is included. Since the measurement of gyrogravitational effect
will be a crucial test of relativistic gravity, theoretical
understanding of normal and anomalous gyrogravitational effects
become important to reveal the significance of the outcome of the
test. Like electromagnetism, Lorentz transformation of electric
field gives rise to magnetic field, while, similarly, "Lorentz
transformation" of gravitoelectric fields result in
gravitomagnetic fields (Lense-Thirring effect) \cite{k,i}. When we
studied the propagation of a Dirac particle in different frames,
in addition to the gravitational effect, the gravitomagnetic
effect also comes out naturally in agreement with previous
investigations.\par In the following subsections of the
introduction, we will review the equation of motion and their
inertial effects for a scalar particle and a Dirac particle. In
sections 2-5, we present new materials and new derivations. In
section 2, the coordinate transformation from a moving accelerated
frame to its instantaneous inertial frame is formulated. In
section 3, we develop a method leading to an exact solution of
Dirac equation in accelerated frames; subsequently, the
gyrogravitational factor of the spin for a Dirac particle will be
found. In section 4, we use eikonal approximation to solve Dirac
equation in accelerated frames. In the end, we have a short
discussion in section 5.\par
\subsection{Coordinate transformation, inertial effects and equation of motion}
\noindent In this section, we review the coordinate transformation
between an inertial observer and an accelerated, rotating
observer, equation of motion in the non-inertial frame, and
associated inertial effects.\par
\subsubsection{The coordinate transformation}                                      
Let an observer who is carrying a tetrad $\{e^\mu(\tau)\}$ with
4-velocity $\bf{u}(\tau)$, 4-acceleration {\bf a}($\tau$) and
4-rotation {\boldmath $\omega$}($\tau$) move along a world line
$\bf{Q}(\tau)$. The observer's local coordinates $\{x^\mu\}$ are
defined by
\begin{equation}
\mbox{\boldmath $\xi$}={x^k}e_k(x^0)+ {\bf Q}(x^0),   \label{b}
\end{equation}
where $\mbox{\boldmath $\xi$}=(\xi^0,\xi^1,\xi^2,\xi^3)$ are the
Minkowski coordinates \cite{a,b,c}. Here and in the following, we
use Greek alphabet to denote 4-indices and Latin alphabet to
denote 3-indices. $b^i$, $\eta^i$, etc. are similarly defined in
terms of ${\bf b}=d{\bf a}/d\tau$, $\mbox{\boldmath
$\eta$}=d\mbox{\boldmath $\omega$}/d\tau$, etc. In the observer's
local coordinates the Minkowski metric is transformed to
\cite{b,c,o}:
\begin{eqnarray}
(ds)^2&=&(d\xi^0)^2-(d\xi^1)^2-(d\xi^2)^2-(d\xi^3)^2,\nonumber\\
 &=&(dx^0)^2[1+2a^ix^i+(a^ix^i)^2-\omega^2(x \cdot x)+(\omega^lx^l)^2]\nonumber\\ & &-2dx^0dx^l\epsilon^{lkm}
\omega^kx^m -dx^ldx^k\delta_{kl}+O[{(\Delta x^\alpha)}^4dx^\beta
dx^\gamma].
\end{eqnarray}
\subsubsection{The equation of motion in local coordinates and inertial effects}             
Consider an inertial motion in the global inertial frame:
\begin{equation}
\xi^i=d^i+v^i\xi^0,\label{d}
\end{equation}
where $v^i$ is the particle's constant 3-velocity and $d^i$ is the
position at $\xi^0=0$. In the local coordinates, the equation of
motion is
\begin{eqnarray}
\frac{d^2x^i}{{dx^0}^2}&=&-a^i(1+a^mx^m)+\frac{w^i+\epsilon^{ijk}\omega^jx^k}{(1+a^mx^m)}[
b^lx^l+2a^lw^l+2\epsilon^{lmn}a^l\omega^mx^n]\nonumber
\\ & & -2\epsilon^{ijk}\omega^jw^k-(\omega^i\omega^m-\omega^2\delta_{im})x^m-\epsilon^{ijk}\eta^jx^k.
\end{eqnarray}
where
\begin{equation}
w^i\,\equiv\,\frac{dx^i}{dx^0}=v^i+(v^ia^l-\epsilon^{ikl}\omega^k)x^l.
\end{equation}
\subsection{The inertial effects of a Dirac particle}In 1990, Hehl
and Ni proposed a framework to study the relativistic inertial
effects of a Dirac particle. The equation of the Dirac particle in
an accelerated and rotating frame is obtained as follows:
\begin{eqnarray}
\hspace*{-.5cm}\{\;i\hbar\;\frac{\gamma^0}{\displaystyle
1+\frac{\vec{a} \cdot \vec{x}}{c^2}}\;[\frac{\partial}{\partial
x^0}\;+ \frac{1}{2c^2}\;\vec{a} \cdot \vec{\alpha}
-\frac{i}{c\hbar}\;\vec{\omega} \cdot \vec{J}\;]+
 \;i\hbar\gamma^i\frac{\partial} {\partial
x^i}\;\}\Psi=mc\;\Psi \label{a4}
\end{eqnarray}
where
\begin{equation}
\vec{J}=\vec{L}+\vec{S}=\vec{x}\times\frac{\hbar}{i}\vec{\frac{\partial}{\partial
x}}+\frac{1}{2}\hbar \;\vec{\sigma}
\end{equation}
is the total angular momentum, $\vec{a}$  is the acceleration, and
$\vec{\omega}$ is the angular velocity. It can be written in the
form
\begin{equation}
i\hbar \frac{\partial}{\partial t}\;\Psi\;=\;H\;\Psi,
\end{equation}
 with the Hamiltonian
\begin{equation}
H=\beta\;mc^2+\mathcal{O}+\mathcal{E},
\end{equation}
where
\begin{eqnarray}
\mathcal{O}&=&c\;\vec{\alpha} \cdot \vec{p} +\frac{1}{2c}[(\vec{a}
\cdot \vec{x})(\vec{p} \cdot \vec{\alpha})+(\vec{p} \cdot
\vec{\alpha})(\vec{a} \cdot \vec{x})], \nonumber\\ \mathcal{E}&=&
\beta m(\vec{a} \cdot \vec{x})-\vec{\omega} \cdot
(\vec{L}+\vec{S}).
\end{eqnarray}
In the non-relativistic approximation, via Foldy-Wouthysen
transformation, the Hamiltonian becomes:
\begin{eqnarray}
H&=&\beta m c^2+\frac{\beta}{2m}p^2+\beta m (\vec{a}\cdot
\vec{x})+ \frac{\beta}{2m}\vec{p}\;[\frac{\vec{a} \cdot \vec{x}}
{c^2}]\cdot \vec{p}\nonumber\\ & &-\vec{\omega} \cdot
(\vec{L}+\vec{S})+\frac{\hbar}{4mc^2}\vec{\sigma}
\cdot(\vec{a}\times \vec{p}).
\end{eqnarray}
This is the Hamiltonian of a Dirac particle in a non-inertial
frame, and a detailed discussion of various terms goes to Hehl \&
Ni \cite{d}. A summary of relativistic inertia effects and their
non-relativistic components of a Dirac particle together with
their empirical significance are given in the table of Hehl \& Ni
\cite{d}.

\section{Moving Accelerated Frame And Lense-Thirring Effect}
\noindent In 1820, Hans Christian Oersted \cite{j} discovered that
electric currents produce a magnetic field. In the mid 19th
century, Maxwell unified the electric field and magnetic field
into the electromagnetic field. As the concept of special
relativity matured, we know that the electric and magnetic fields
are just one thing interpreted by different observers. According
to Einstein's general relativity, the currents of mass can produce
a field which is called, by analogy, the gravitomagnetic field
\cite{k,l,m}.\par In this section, we consider the gravitomagnetic
effects of a scalar particle. A rotation of the coordinate due to
the observer's velocity with respect to the source of the gravity
will be found through coordinate transformations. The results of
this section will be used for the following discussion.\par

Like the way in Section 1, a coordinate transformation should be
found first for solving the equation. However, it is not readily
available to accquire a coordinate transformation between an
accelerated frame and another accelerated frame moving with
respect to the gravity source. Even though the velocity between
them is constant, Lorentz transformation fails. The reason is that
both observers in the moving frame and in the rest frame of the
source are not in the inertial frames. Nevertheless, since
$\vec{a}=0$, the coordinate transformation between them is still
Lorentz transformation. It is hoped that the Lorentz
transformation will be approximately true.\par Here we will
construct the relations among four coordinate frames. Suppose
there is an inertial frame $\{\xi^\mu\}$. At $\xi^0=0$, another
frame $\{x^\mu\}$ whose origin coincides that of the inertial
frame $\{\xi^\mu\}$ is accelerated uniformly in the
$\xi^1$-direction, and the acceleration is $\vec{a}=(a,0,0)$.
According to MTW \cite{a}, the coordinate transformation is
\begin{eqnarray}
\xi^0 &=& (\frac{c^2}{a}+x^1)\sinh\frac{ax^0}{c^2} \nonumber\\
\xi^1 &=& (\frac{c^2}{a}+x^1)\cosh\frac{ax^0}{c^2}-\frac{c^2}{a}
\nonumber\\ \xi^2 &=& x^2 \nonumber\\ \xi^3 &=& x^3.\label{uu}
\end{eqnarray}
This transformation is the coordinate transformation which we used
in the last section. At the same time, there is another
accelerated frame $\{y^\mu\}$ which moves with the velocity
$\vec{w}$=$(0,w,0)$ with respect to the uniformly accelerated
frame $\{x^\mu\}$ where the gravity can be viewed as static.
Though, the transformation between $\{x^\mu\}$ and $\{y^\mu\}$ is
not known exactly, the transformation should approximately be
Lorentz transformation:
\begin{eqnarray}
x^0&\cong&(1+\frac{w^2}{2c^2})y^0-\frac{w}{c}(1+\frac{w^2}{2c^2})y^2\nonumber\\
x^1&\cong&y^1\nonumber\\
x^2&\cong&(1+\frac{w^2}{2c^2})y^2-\frac{w}{c}(1+\frac{w^2}{2c^2})y^0\nonumber\\
x^3&\cong&y^3.\label{u}
\end{eqnarray}
At $y^0=0$, there exists an inertial frame $\{\zeta^\mu\}$ whose
velocity with respect to the moving accelerated frame $\{y^\mu\}$
vanishes. The coordinate transformation between them should be in
the standard form stated in Li \& Ni \cite{b}. since at
$x^0=\xi^0=0$ the velocity between the frames, $\{x^\mu\}$ and
$\{\xi^\mu\}$, is zero, the inertial frame $\{\zeta^\mu\}$ also
travels with the velocity $\vec{w}$=$(0,w,0)$ with respect to the
other inertial frame $\{\xi^\mu\}$. Because both $\{\xi^\mu\}$ and
$\{\zeta^\mu\}$ are inertial frames, their transformations, of
course, are Lorentz transformations. \par By successive coordinate
transformation between $\{\zeta^\mu\}$ and $\{\xi^\mu\}$,
$\{\xi^\mu\}$ and $\{x^\mu\}$, $\{x^\mu\}$ and $\{y^\mu\}$, we can
obtain the approximate transformation from $\{y^\mu\}$ to
$\{\zeta^\mu\}$ (the validity of the transformation (\ref{gg}) is
discussed in Appendix A):
\begin{eqnarray}
\zeta^0&\cong&
y^0+\frac{a}{c^2}y^1y^0+\frac{a^2}{6c^4}{y^0}^3\nonumber\\
\zeta^1&\cong&
y^1+\frac{a}{2c^2}(1+\frac{w^2}{c^2}){y^0}^2+\frac{aw}{c^3}y^2y^0+\frac{a^2}{2c^4}y^1{y^0}^2\nonumber\\
\zeta^2&\cong& y^2-\frac{aw}{c^3}y^1y^0\nonumber\\
\zeta^3&=&y^3.\label{gg}
\end{eqnarray}
By comparing (\ref{gg}) with the work of Li \& Ni \cite{b}, we
conclude that in the moving accelerated frame $\{y^\mu\}$ the
acceleration $\vec{a}^\prime$ is:
\begin{equation}
\vec{a}^\prime\cong(1+\frac{w^2}{c^2})\vec{a}
\end{equation}
and a rotation $\vec{\omega}$ will be produced:
\begin{equation}
\vec{\omega}=(0,0,-\frac{aw}{c^2}).
\end{equation}
These are the gravitomagnetic effects which result from the motion
of the gravity source.\par
\section{PROPAGATION OF DIRAC WAVE FUNCTIONS IN ACCELERATED FRAMES}
\subsection{Searching for the transformation operator}
In this section, a different approach to an exact solution of the
non-inertial Dirac equation will be performed in contrast to the
approximate method stated in the last section. In the following,
we try to find a transformation of a Dirac particle from an
inertial frame to an accelerated frame. As the usual solution of
Dirac equation in inertial frame is known, we can operate the
transformation operator on the solution in the inertial frame to
obtain the exact solution in the accelerated frame.\par Dirac
equation in the inertial frame $(\xi^0,\xi^1,\xi^2,\xi^3)$ is:
\begin{equation}
i\hbar\{\gamma^0\frac{\partial}{\partial
\xi^0}+\gamma^1\frac{\partial}{\partial \xi^1}
+\gamma^2\frac{\partial}{\partial
\xi^2}+\gamma^3\frac{\partial}{\partial \xi^3}\} \Psi^\prime =
mc\; \Psi^\prime, \label{l}
\end{equation}
while Dirac equation in the uniformly accelerated frame
$(x^0,x^1,x^2,x^3)$ is:
\begin{eqnarray}
\hspace*{-1cm}i\hbar\{\frac{\gamma^0}{1+\frac{ax^1}{c^2}}\frac{\partial}{\partial
x^0}+\frac{a/2c^2}{1+
\frac{ax^1}{c^2}}\gamma^1+\gamma^1\frac{\partial}{\partial
x^1}+\gamma^2\frac{\partial}{\partial
x^2}+\gamma^3\frac{\partial}{\partial x^3}\}\;\Psi=
\frac{mc}{\hbar}\; \Psi. \label{m}
\end{eqnarray}
The coordinate transformation between the inertial frame and the
uniformly accelerated frame is:
\begin{eqnarray}
\xi^0 &=& (\frac{c^2}{a}+x^1)\sinh\frac{ax^0}{c^2} \nonumber\\
\xi^1 &=& (\frac{c^2}{a}+x^1)\cosh\frac{ax^0}{c^2}-\frac{c^2}{a}
\nonumber\\ \xi^2 &=& x^2 \nonumber\\ \xi^3 &=& x^3 \label{n}
\end{eqnarray}
If we assume a transformation {\bf S}, which is the matrix
function of $x^0$ and $x^1$ only, exists, i.e.
\begin{equation}
\Psi^\prime = {\bf S}\;\Psi,
\end{equation}
then ${\bf S}$ should satisfy the conditions:
\begin{eqnarray}
{\bf
S}^{-1}(\cosh\frac{ax^0}{c^2}\;\gamma^0-\sinh\frac{ax^0}{c^2}\;\gamma^1){\bf
S}=\gamma^0,\hspace*{1cm}\label{o}\\ {\bf
S}^{-1}(\cosh\frac{ax^0}{c^2}\;\gamma^1-\sinh\frac{ax^0}{c^2}\;\gamma^0){\bf
S}=\gamma^1,\hspace*{1cm}\label{p}\\
\frac{1}{1+\frac{ax^1}{c^2}}\gamma^0{\bf
S}^{-1}\frac{\partial}{\partial x^0}{\bf S}+\gamma^1{\bf S}^{-1}
\frac{\partial}{\partial x^1}{\bf
S}=\frac{a/2c^2}{1+\frac{ax^1}{c^2}}\gamma^1\label{q},
\end{eqnarray}
\begin{eqnarray}
{\bf S}^{-1}\gamma^2{\bf S}&=&\gamma^2,\label{r}\\ {\bf
S}^{-1}\gamma^3{\bf S}&=&\gamma^3.\label{s}
\end{eqnarray}
From the equations ({\ref o}) and ({\ref p}), we suppose that {\bf
S} is in the form:
\begin{equation}
{\bf S}=\exp(\varrho\;\gamma^0\gamma^1).\label{2}
\end{equation}
Substituting the assumption (\ref{2}) into the equations ({\ref
o}) and (\ref{p}), then we can find that
\begin{equation}
\varrho=\frac{a}{2c^2}x^0,
\end{equation}
and the equation ({\ref q}) will be satisfied.\par Since
$\gamma^2$ and $\gamma^3$ are commute with $\gamma^0\gamma^1$, the
conditions (\ref{r}) and (\ref{s}) are satisfied. So the equation
(\ref{l}) can become the equation (\ref{m}) by the coordinate
transformation and the transformation operator
\begin{equation}
{\bf S}=\;\exp\,(\;\frac{ax^0}{2c^2}\;\alpha_1\;).
\label{operator}
\end{equation}
In the inertial frame, an electron with the polarization in the
$x^3$-direction can be described by
\begin{equation}
\Psi^\prime=\frac{1}{\sqrt{\frac{2mc}{\hbar}(k^{(0)}_0+\frac{mc}{\hbar})}}\left[
\begin{array}{c} k^{(0)}_0+\frac{mc}{\hbar}\\
0\\0\\-k^{(0)}_+
\end{array}
\right] e^{-i(k^{(0)}_0\xi^0+k^{(0)}_1\xi^1+k^{(0)}_2\xi^2)},
\end{equation}
with the constant wave vector
$k^{(0)}_\mu=(k^{(0)}_0,k^{(0)}_1,k^{(0)}_2,0)$. As we expected in
the beginning, the wave function can be obtained by operating
${\bf S}^{-1}$ on $\Psi^\prime$:
\begin{equation}
\Psi\;\;=\;\;{\bf S}^{-1}\;\Psi^\prime.
\end{equation}
Hence the exact solution is
\begin{eqnarray}
\hspace*{-.5cm}\Psi=\frac{e^{-i\theta}}{\sqrt{\frac{2mc}{\hbar}(k^{(0)}_0+\frac{mc}{\hbar})}}\left[\begin{array}{c}\displaystyle
(k^{(0)}_0+\frac{mc}{\hbar})\cosh\frac{ax^0}{2c^2}+k^{(0)}_+\sinh\frac{ax^0}{2c^2}\\
0\\0\\ \displaystyle
-(k^{(0)}_0+\frac{mc}{\hbar})\sinh\frac{ax^0}{2c^2}-k^{(0)}_+\cosh\frac{ax^0}{2c^2}
\end{array}\right].
\end{eqnarray}
When $x^0$=$0$, {\bf S} is the identity matrix, the initial wave
vector in the accelerated frame is also
 $k_\mu(x^0=0)$=$(k^{(0)}_0,k^{(0)}_1,k^{(0)}_2,0)$. As to the phase, because $\theta$ is a scalar, we can
get the phase presented in the accelerated frame through
coordinate transformation (\ref{n}):
\begin{eqnarray}
\hspace*{-1cm}\theta=k^{(0)}_0(\frac{c^2}{a}+x^1)\sinh\frac{ax^0}{c^2}+k^{(0)}_1(\frac{c^2}{a}+x^1)\cosh\frac{ax^0}{c^2}
+k^{(0)}_2x^2-k^{(0)}_1\frac{c^2}{a}.
\end{eqnarray}
The definition of a wave vector is
\begin{equation}
k_\mu=\frac{\partial}{\partial x^\mu}\;\theta,
\end{equation}
so the wave vector in the accelerated frame reads:
\begin{eqnarray}
k_0 &=&
k^{(0)}_0(1+\frac{ax^1}{c^2})\cosh\frac{ax^0}{c^2}+k^{(0)}_1(1+\frac{ax^1}{c^2})\sinh\frac{ax^0}{c^2},\nonumber\\
k_1 &=&
k^{(0)}_0\sinh\frac{ax^0}{c^2}+k^{(0)}_1\cosh\frac{ax^0}{c^2},\nonumber\\
k_2 &=& k^{(0)}_2,\nonumber\\ k_3 &=& 0.
\end{eqnarray}
Consider a Dirac particle with the polarization in the $x^1$-$x^2$
plane thrown at $x^0=0$.
\begin{eqnarray}
\Psi = \frac{\sqrt{2}/2\times
e^{-i\theta}}{\sqrt{\frac{2mc}{\hbar}(k^{(0)}_0+\frac{mc}{\hbar})}}\;
\left[\begin{array}{cccc} e^{\displaystyle
-i\frac{\phi}{2}}((k^{(0)}_0+\frac{mc}{\hbar})\cosh\frac{ax^0}{2c^2}+k^{(0)}_+\sinh\frac{ax^0}{2c^2})\\
e^{\displaystyle
i\frac{\phi}{2}}((k^{(0)}_0+\frac{mc}{\hbar})\cosh\frac{ax^0}{2c^2}+k^{(0)}_\_\sinh\frac{ax^0}{2c^2})\\
-e^{\displaystyle
i\frac{\phi}{2}}(k^{(0)}_\_\cosh\frac{ax^0}{2c^2}+(k^{(0)}_0+\frac{mc}{\hbar})\sinh\frac{ax^0}{2c^2})\\
-e^{\displaystyle
-i\frac{\phi}{2}}(k^{(0)}_+\cosh\frac{ax^0}{2c^2}+(k^{(0)}_0+\frac{mc}{\hbar})\sinh\frac{ax^0}{2c^2})
\end{array}\right],\nonumber\\
\end{eqnarray}
where $\phi$ is the angle between $x^1$-axis and the particle's
polarization. When the particle hits the floor of the accelerated
frame, in the movement of the particle with the velocity
$(v^{(0)}_{x^1},v^{(0)}_{x^2},0)$ at $x^0=0$ in the uniformly
accelerated frame, the movement should be symmetric in time. It
means that when the particle falls back to the floor, its velocity
must be $(-v^{(0)}_{x^1},v^{(0)}_{x^2},0)$. Therefore
\begin{equation}
k_1|_{when\;the\;particle\;hits\;the\;floor}=-k^{(0)}_1.
\end{equation}\par
It will give
\begin{equation}
x^0= \frac{c^2}{a}\tanh^{-1}\frac{2v^{(0)}_{x^1}/c}{\displaystyle
1+\frac{{v^{(0)}_{x^1}}^2}{c^2}}.
\end{equation}\par
This is the time of a flying rod coming back to the floor.\par
\subsection{The uniqueness problem}
Suppose there exists another matrix ${\bf \bar{S}}$, which also
satisfies the conditions (\ref{o})$\sim$(\ref{s}), from (\ref{o})
and (\ref{p}) with ${\bf S}$ replaced by ${\bf \bar{S}}$, we
derive
\begin{eqnarray}
\gamma^0 &=& {\bf
\bar{S}}(\cosh\frac{ax^0}{c^2}\gamma^0+\sinh\frac{ax^0}{c^2}\gamma^1){\bf
\bar{S}}^{-1}\label{at}\\ \gamma^1 &=& {\bf
\bar{S}}(\cosh\frac{ax^0}{c^2}\gamma^1+\sinh\frac{ax^0}{c^2}\gamma^0){\bf
\bar{S}}^{-1}.\label{au}
\end{eqnarray}
If we substitute (\ref{at}) and (\ref{au}) into the left hand side
of (\ref{o}) and (\ref{p}), we obtain
\begin{eqnarray}
({\bf \bar{S}}^{-1}{\bf S})^{-1}\gamma^0{\bf \bar{S}}^{-1}{\bf S}
&=& \gamma^0\label{ba}\\ ({\bf \bar{S}}^{-1}{\bf
S})^{-1}\gamma^1{\bf \bar{S}}^{-1}{\bf S} &=& \gamma^1.\label{bb}
\end{eqnarray}
Similarly,
\begin{eqnarray}
({\bf \bar{S}}^{-1}{\bf S})^{-1}\gamma^2{\bf \bar{S}}^{-1}{\bf S}
&=& \gamma^2\label{bc}\\ ({\bf \bar{S}}^{-1}{\bf
S})^{-1}\gamma^3{\bf \bar{S}}^{-1}{\bf S} &=& \gamma^3.\label{bd}
\end{eqnarray}
Since ${\bf \bar{S}}^{-1}{\bf S}$ commutes with $\gamma^\mu$
$(\mu=0\sim3)$ , it must be a constant matrix and ${\bf \bar{S}}$
is equal to ${\bf S}$ up to a constant scale factor.\par
\subsection{Gyrogravitational factor for the spin of a Dirac particle}
\subsubsection{The polarization of a Dirac particle in the moving accelerated frame}
According to the previous section, we know that a scalar particle
in the moving accelerated frame will experience a Coriollis force
due to the rotation of the frame. In this section, we consider how
the polarization of a Dirac particle changes, when it is placed in
the moving accelerated frame.\par At $y^0=0$, the frame
$\{\zeta^\mu\}$ is at rest for the observers in $\{y^\mu\}$.
Therefore, the polarization of the Dirac particle in
$\{\zeta^\mu\}$ is a constant vector. We can use the Lorentz
transformation and the transformation operator {\bf S} to obtain
the Dirac wave function in the moving accelerated frame. In the
similar way stated in th last part of this section, we use the
operator:
\begin{equation}
{\bf L}(w)\;{\bf S}^{-1}(a)\;{\bf L}(-w),
\end{equation}
where ${\bf L}(w)$ is the Lorentz transformation:
\begin{eqnarray}
{\bf
L}(w)=\frac{1}{\sqrt{\frac{2mc}{\hbar}(k^w_0+\frac{mc}{\hbar})}}
\left[ \begin{array}{c} k^w_0+\frac{mc}{\hbar}\\ 0\\ 0\\
ik^w_2\end{array}
\begin{array}{c} 0\\k^w_0+\frac{mc}{\hbar}\\-ik^w_2\\0
\end{array}\begin{array}{c} 0\\ik^w_2\\k^w_0+\frac{mc}{\hbar}\\0
\end{array}\begin{array}{c} -ik^w_2\\0\\0\\k^w_0+\frac{mc}{\hbar}
\end{array}\right],\nonumber\\
\end{eqnarray}
with
\begin{equation}
k^w_2=-\frac{\frac{mc}{\hbar}\frac{w}{c}}{\sqrt{1-\frac{w^2}{c^2}}},\hspace*{2cm}k^w_0=\frac{\frac{mc}{\hbar}}{\sqrt{1-\frac{w^2}{c^2}}}.
\end{equation}
For the first order of $a$, we find
\begin{eqnarray}
&&{\bf L}(w)\;{\bf S}^{-1}(a)\;{\bf L}(-w)\cong\nonumber\\&&
\left[
\begin{array}{c} 1+i\frac{aw}{2c^3}y^0\\ 0\\ 0\\
-(1+\frac{w^2}{c^2})\frac{ay^0}{2c^2}\end{array}
\begin{array}{c} 0\\ 1-i\frac{aw}{2c^3}y^0\\-(1+\frac{w^2}{c^2})\frac{ay^0}{2c^2}\\0
\end{array} \begin{array}{c} 0\\-(1+\frac{w^2}{c^2})\frac{ay^0}{2c^2}\\1+i\frac{aw}{2c^3}y^0\\0
\end{array} \begin{array}{c} -(1+\frac{w^2}{c^2})\frac{ay^0}{2c^2}\\0\\0\\1-i\frac{aw}{2c^3}y^0
\end{array}\right].\nonumber\\
\end{eqnarray}
Now the operator can be denoted as
\begin{equation}
{\mathcal S}^{-1}(a^\prime,\omega_3)={\bf L}(w)\;{\bf
S}^{-1}(a)\;{\bf L}(-w).
\end{equation}
Just taking the linear part of ${\mathcal
S}^{-1}(a^\prime,\frac{\omega_3}{2})$ into account, it can divided
into
\begin{equation}
{\mathcal S}^{-1}(a^\prime,\omega_3)\cong{\bf
S}^{-1}(a^\prime){\bf R}(\frac{\omega_3y^0}{c}),
\end{equation}
where
\begin{equation}
{\bf S}^{-1}(a^\prime)\cong \left[ \begin{array}{c} 1\\ 0\\ 0\\
-\frac{a'y^0}{2c^2}\end{array}
\begin{array}{c} 0\\ 1\\-\frac{a'y^0}{2c^2}\\0
\end{array}\begin{array}{c} 0\\-\frac{a'y^0}{2c^2}\\1\\0
\end{array}\begin{array}{c}
-\frac{a'y^0}{2c^2}\\0\\0\\1
\end{array}\right]
\end{equation}
and
\begin{equation}
{\bf R}(\frac{\omega_3y^0}{c})\cong \left[ \begin{array}{c}
1+i\frac{aw}{2c^3}y^0\\ 0\\ 0\\ 0\end{array}
\begin{array}{c} 0\\1-i\frac{aw}{2c^3}y^0\\0\\0
\end{array}\begin{array}{c} 0\\0\\1+i\frac{aw}{2c^3}y^0\\0
\end{array}\begin{array}{c} 0\\0\\0\\1-i\frac{aw}{2c^3}y^0
\end{array}\right].
\end{equation}
${\bf S}^{-1}(a^\prime)$ is the first order approximation of the
operator ${\bf S}^{-1}$, and ${\bf R}(\frac{\omega_3y^0}{c})$ is
the first approximation of the rotation operator:
\begin{eqnarray}
{\bf R}(\frac{\omega_3y^0}{c})&\cong&
I-i\sigma_3\frac{\omega_3}{2c}y^0\\
&\cong&\exp(-i\sigma_3\frac{\omega_3}{2c}y^0).
\end{eqnarray}
Therefore, we can find the gyrogravitational factor is $1$.\par
\subsubsection{Deriving Dirac equation in the moving accelerated frame}
In section 1, a scalar particle's equation of motion can be
obtained, after the coordinate transformation is known. As to
Dirac equation, an additional transformation operator, like {\bf
S} and {\bf L}, is needed, when we derive the equation in some
frames. In this subsection, we want to figure out Dirac equation
in the moving accelerated frame. Since we have the operator
$\mathcal{S}$ and the coordinate transformation (\ref{gg}), the
Dirac equation in the inertial frame $\{\zeta^\mu\}$ is:
\begin{equation}
i\hbar\{\gamma^0\frac{\partial}{\partial
\zeta^0}+\gamma^1\frac{\partial}{\partial \zeta^1}
+\gamma^2\frac{\partial}{\partial
\zeta^2}+\gamma^3\frac{\partial}{\partial \zeta^3}\} \Psi^\prime =
mc\; \Psi^\prime.\label{a1}
\end{equation}
Suppose $\Psi$ is Dirac wave function in the moving accelerated
frame $\{y^\mu\}$. The relation between these functions are:
\begin{eqnarray}
\Psi \cong {\mathcal{S}}^{-1}(a',\omega_3)\Psi'\label{a2}
\end{eqnarray}
Using chain rule and substituting (\ref{a2}) into (\ref{a1}), we
get
\begin{eqnarray}
\hspace*{-.5cm}i\hbar{\mathcal{S}}^{-1}\{\gamma^0\frac{\partial
y^\mu}{\partial \zeta^0}\frac{\partial}{\partial
y^\mu}+\gamma^1\frac{\partial y^\mu}{\partial
\zeta^1}\frac{\partial}{\partial y^\mu} +\gamma^2\frac{\partial
y^\mu}{\partial \zeta^2}\frac{\partial}{\partial
y^\mu}+\gamma^3\frac{\partial y^\mu}{\partial
\zeta^3}\frac{\partial}{\partial
y^\mu}\}{\mathcal{S}}\Psi=mc\Psi.\nonumber\\
\end{eqnarray}
Then it becomes
\begin{eqnarray}
i\hbar\hspace*{0cm}\{\hspace*{0cm}(1-\frac{a'}{c^2}y^1+\frac{{a'}^2}{2c^4}{y^0}^2)\gamma^0\frac{\partial}{\partial
y^0}+\frac{a'}{2c^2}\gamma^1+(1+\frac{{a'}^2}{2c^4}{y^0}^2)\gamma^1\frac{\partial}{\partial
y^1}+\gamma^2\frac{\partial}{\partial
y^2}\hspace{1cm}\nonumber\\+\gamma^3\frac{\partial}{\partial y^3}
+i\frac{a'w}{2c^3}\sigma^3+\frac{a'w}{c^3}(y^1\frac{\partial}{\partial
y^2}-y^2\frac{\partial}{\partial
y^1})+(\frac{a'}{c^4}y^1y^0+\frac{{a'}^3}{6c^6}{y^0}^3)\gamma^0\frac{\partial}{\partial
y^1}\nonumber\\-\frac{{a'}^2w}{c^5}{y^0}^2\gamma^0\frac{\partial}{\partial
y^2} +\frac{a'w}{c^3}y^0(\gamma^1\frac{\partial}{\partial
y^2}-\gamma^2\frac{\partial}{\partial
y^1})\}\Psi\;\;=\;\;mc\,\Psi.
\end{eqnarray}
Setting $y^0$=0, then we obtain
\begin{eqnarray}
i\hbar\{(1-\frac{a'}{c^2}y^1)\gamma^0\frac{\partial}{\partial
y^0}+\frac{a'}{2c^2}\gamma^1+\gamma^1\frac{\partial}{\partial
y^1}+\gamma^2\frac{\partial}{\partial
y^2}+\gamma^3\frac{\partial}{\partial y^3}
\hspace*{1.5cm}\nonumber\\+i\frac{a'w}{2c^3}\sigma^3+\frac{a'w}{c^3}(y^1\frac{\partial}{\partial
y^2}-y^2\frac{\partial}{\partial y^1})\}\Psi\;\;=\;\;mc\,\Psi.
\end{eqnarray}
Rewrite it as
\begin{eqnarray}
i\hbar\{(1-\frac{a'}{c^2}y^1)\gamma^0\frac{\partial}{\partial
y^0}+\frac{a'}{2c^2}\gamma^1-\frac{i}{c\hbar}\omega_3(\frac{\hbar}{2}\sigma^3+y^1\frac{\hbar}{i}\frac{\partial}{\partial
y^2}-y^2\frac{\hbar}{i}\frac{\partial}{\partial
y^1})\hspace{1cm}\nonumber\\ +\gamma^1\frac{\partial}{\partial
y^1}+\gamma^2\frac{\partial}{\partial
y^2}+\gamma^3\frac{\partial}{\partial y^3}\}\Psi\;\;=\;\;mc\,\Psi.
\label{a3}\end{eqnarray} This is familiar to us, since (\ref{a3})
is the approximate form of non-inertial Dirac equation
(\ref{a4}).\par
\section{EIKONAL APPROXIMATION}
\subsection{Eikonal approximation method}                                                
The behavior of a rod thrown in an accelerated frame has been
investigated in detail by K. Nordtvedt \cite{n}. We will extend
his work to the problem of a Dirac particle observed in a
uniformly accelerated frame. To develop an approximate method for
solving the equation is reasonably a good start point to
understand it.\par The wave vector can be defined by
\begin{equation}
 k_\mu =\frac{\partial}{\partial x^\mu}\;\theta,   \label{1}
\end{equation}
where $\theta$, the phase of the wave function, is a scalar. It is
easy to obtain the wave vector in any coordinates through
coordinate transformation. The required coordinate transformation
between a global inertial frame and a local non-inertial frame is
well-defined in Li \& Ni \cite{b} and the coordinate
transformation from an inertial frame to a uniformly accelerated
frame is known as (\ref{uu}).\par If the wave vector in the
inertial frame is {\bf k}$=(k^{(0)}_0,k^{(0)}_1,k^{(0)}_2,0)$,
substitute (\ref{uu}) into the phase $\theta$, where
$\theta=k^{(0)}_\mu \xi^\mu$ in the inertial frame. Therefore, the
terms of the phase $\theta$ can be classified into some groups
according to their order of $c$.
\begin{eqnarray}
\theta^{(0)}&=&k^{(0)}_0x^0+k^{(0)}_1x^1+k^{(0)}_2x^2 \nonumber\\
\theta^{(1)}&=&\frac{a}{2c^2}k^{(0)}_1(x^0)^2+\frac{a}{c^2}k^{(0)}_0x^0x^1
\nonumber\\ \theta^{(2)}&=&\frac{a^2}{6c^4}\;k^{(0)}_0
(x^0)^3+\frac{a^2}{2c^4}k^{(0)}_1x^1{(x^0)}^2\nonumber\\ &
&\vdots\label{111}
\end{eqnarray}
After (\ref{111}) is taken into the equation (\ref{1}), the wave
vector in the uniformly accelerated frame is obtained in an
approximate form:
\begin{eqnarray}
k_0 &\cong&
k^{(0)}_0+\frac{a}{c^2}k^{(0)}_1x^0+\frac{a}{c^2}k^{(0)}_0x^1+\nonumber\\
&&\frac{a^2}{2c^4}k^{(0)}_0{x^0}^2+\frac{a^2}{c^4}k^{(0)}_1x^0x^1+\frac{a^3}{2c^6}k^{(0)}_0x^1{x^0}^2\nonumber\\
k_1 &\cong&
k^{(0)}_1+\frac{a}{c^2}k^{(0)}_0x^0+\frac{a^2}{2c^4}k^{(0)}_1{x^0}^2+\frac{a^3}{6c^6}k^{(0)}_0{x^0}^3\nonumber\\
k_2 &\cong& k^{(0)}_2 \nonumber \\ k_3 &=& 0. \label{k}
\end{eqnarray}\par


A Dirac particle with a certain polarization, at $x^0=0,$ is
placed at the origin of the accelerated coordinate, which is
boosted in the direction of $x^1$. Without comprising any
generality, in simplicity, we suppose the particle has an initial
velocity $\vec{v}(x^0=0)\;=\;(v^{(0)}_{x^1},v^{0)}_{x^2},0)$.
Dirac equation in a uniformly accelerated frame presented in the
last section can be rewritten as:
\begin{eqnarray}
i\hbar\{\;\gamma^0\frac{\partial}{\partial
x^0}+(1+\varepsilon\frac{ax^1}{c^2})\gamma^i\frac{\partial}{\partial
 x^i}+\varepsilon\frac{a}{2c^2}\gamma^1\}\Psi\;=m\,c(1+\varepsilon\frac{ax^1}{c^2})\;\Psi, \label{f}
\end{eqnarray}
we assume $\Psi$ is in the form
\begin{equation}
\Psi=(\;\Psi^{(0)}+\varepsilon\;\Psi^{(1)}\;
+\varepsilon^2\;\Psi^{(2)} +\cdots\;)\;e^{-i\,\theta}\label{h},
\end{equation}
where $\theta$ is the phase of the Dirac wave-function and
$\Psi^{(0)}$ is a constant 4-component column vector, $\Psi^{(1)}$
is a 4-component column vector function of the 1st order $1/c^2$
of $x^0$,$x^1$, $\Psi^{(2)}$ is a 4-component column vector
function of the 2nd order $1/c^2$ of $x^0$,$x^1$, $\cdots$ and so
on. Herein, $\varepsilon$ is the parameter to keep track of the
order of $1/c^2$, and it will be set 1 after solving the
equation.\par Substituting (\ref{h}) into (\ref{f}), $\Psi^{(0)}$
$\Psi^{(1)}$ \ldots can be all solved order by order. For the
order of 1, the equation reads
\begin{equation}
\gamma^0\;\;\theta^{(0)}_{,0}\;\Psi^{(0)}\;\;+\;\;\gamma^i\;\;\theta^{(0)}_{,i}\;\Psi^{(0)}\;=\;
\frac{mc}{\hbar}\;\Psi^{(0)}.\label{i}
\end{equation}
This equation for $\Psi^{(0)}$ is the usual form of the Dirac
equation in an inertial frame. According to the wave vector that
was obtained in the beginning of this section,
\begin{equation}
\theta^{(0)}_{,0}=k^{(0)}_0=const.
\;\;,\;\;\theta^{(0)}_{,i}=k^{(0)}_i=const.
\end{equation}
For the first order of $\varepsilon$, we have
\begin{eqnarray}
i\gamma^0\Psi^{(1)}_{,0}+i\gamma^i\Psi^{(1)}_{,i}+(\gamma^0k^{(0)}_0+
\gamma^ik^{(0)}_i-\frac{mc}{\hbar})\Psi^{(1)}\hspace*{4cm}\nonumber\\
=-(\gamma^0\theta^{(1)}_{,0}+\gamma^i\theta^{(1)}_{,i})\;\Psi^{(0)}
-\frac{ax^1}{c^2}(\gamma^ik^{(0)}_i-\frac{mc}{\hbar})\Psi^{(0)}-i\frac{a}{c^2}\gamma^1
\Psi^{(0)}.\label{j}
\end{eqnarray}
Now the equation (\ref{i}) and (\ref{j}) will be used to solve
$\Psi^{(0)}$ and $\Psi^{(1)}$ with two initial conditions as
examples:
\begin{itemize}
\item EXAMPLE 1\\
In the global inertial frame \{$\xi^\mu$\}, if the particle has
the constant velocity
$\vec{v}(x^0$=$0)$=$(v^{(0)}_{x^1},v^{(0)}_{x^2},0)$ and the
polarization oriented in the $x^3$-direction, the wave function
is:
\begin{equation}
\Psi^\prime|_{\xi^0=0}=\frac{1}{\sqrt{\frac{2mc}{\hbar}(k^{(0)}_0+\frac{mc}{\hbar})}}
\left[\begin{array}{cccc}
k^{(0)}_0+\frac{mc}{\hbar}\\0\\0\\k^{(0)}_+
\end{array}\right]
\end{equation}
where $k^{(0)}_+\equiv\;k^{(0)}_1+ik^{(0)}_2$ (in the similar
way,$\;$ $k^{(0)}_\_\equiv\;k^{(0)}_1-ik^{(0)}_2$). At
$\xi^0=x^0=0$, since the relative velocity between two frames
vanishes and from the equation (\ref{i}), we know
\begin{equation}
\Psi^\prime|_{\xi^0=0}=\Psi|_{x^0=0}=\Psi^{(0)}.
\end{equation}
$\Psi^{(1)}$ can be obtained by solving (\ref{j}) with a simple
assumption. Consequently, we get the approximate solution:
\begin{eqnarray}
\Psi &\cong&
(\Psi^{(0)}+\Psi^{(1)})e^{i(\theta^{(0)}+\theta^{(1)})}\hspace*{3cm}\nonumber\\
&=&\frac{1}{\sqrt{\frac{2mc}{\hbar}(k^{(0)}_0+\frac{mc}{\hbar})}}
\left[\begin{array}{cccc}
k^{(0)}_0+\frac{mc}{\hbar}+\frac{ax^0}{2c^2}k^{(0)}_+\\ 0\\0\\
-k^{(0)}_+-\frac{ax^0}{2c^2}(k^{(0)}_0+\frac{mc}{\hbar})
\end{array}\right]\nonumber\\& &\hspace*{2cm}\times e^{[-i(k^{(0)}_0x^0+k^{(0)}_1x^1+k^{(0)}_2x^2+\frac{a}{2c^2}k^{(0)}_1
{x^0}^2+\frac{a}{c^2}k^{(0)}_0x^1x^0)]} \nonumber\\
\end{eqnarray}
At the moment $x^0=\displaystyle
\frac{c^2}{a}\tanh^{-1}\frac{2v^{(0)}_{x^1}/c}{1+\frac{{v^{(0)}_{x^1}}^2}{c^2}}
$, when the particle hits the floor of the accelerated frame, the
speed of the particle is
$\vec{v}(x^0=0)=(-v^{(0)}_{x^1},v^{(0)}_{x^2},0)$ and its wave
vector is  $k_\mu= (k^{(0)}_0,-k^{(0)}_1,k^{(0)}_2,0)$. We can
transform $\Psi$ to its rest frame to identify the polarization of
the particle via Lorentz transformation {\bf L}. Lorentz
transformation of a Dirac particle is
\begin{eqnarray}
{\bf L}=exp(\displaystyle \frac{\omega}{2}\frac{\vec{\alpha} \cdot
\vec{v}}{|\vec{v}|}),
\end{eqnarray}
where $\vec{v}=(-v^{(0)}_{x^1},v^{(0)}_{x^2},0)$,
$cosh\frac{\omega}{2}=\sqrt{\frac{ E+mc^2}{2mc^2}}$. Applying
Lorentz transformation to the wave function in the accelerated
frame, we get
\begin{eqnarray}
\Psi|_{in\;particle's\;rest\;frame}={\bf
L}\;\Psi|_{in\;accelerated\;frame}\hspace*{4cm}\nonumber\\ \cong
\frac{1}{\sqrt{\frac{2mc}{\hbar}(k^{(0)}_0+\frac{mc}{\hbar})}}
\left[\begin{array}{c}
{(k^{(0)}_0+\frac{mc}{\hbar})}^2+{(k^{(0)}_+)}^2+\frac{ax^0}{c^2}k^{(0)}_+
(k^{(0)}_0+\frac{mc}{\hbar})\\ 0\\ 0\\ 0 \end{array}
\right]\nonumber\\ \hspace*{0cm}\times
e^{-i(\theta^{(0)}+\theta^{(1)}+\cdots)}|_{x^0=\frac{c^2}{a}\tanh^{-1}\frac{2v^{(0)}_{x^1}/c}{1+\frac{{v^{(0)}_{x^1}}^2}{c^2}}}.
\end{eqnarray}
Obviously, its polarization of spin doesn't change at all.
\item EXAMPLE 2 \\
In this example, a Dirac particle with the polarization in the
$x^1$-$x^2$ plane is thrown in the accelerated frame at $x^0=0$.
At this moment, the wave function can be written as follows:
\begin{equation}
\Psi|_{x^0=0}=\frac{\sqrt2/2}{\sqrt{\frac{2mc}{\hbar}(k^{(0)}_0+\frac{mc}{\hbar})}}
\left[\begin{array}{cccc} e^{\displaystyle
-i\frac{\phi}{2}}(k^{(0)}_0+\frac{mc}{\hbar})\\ e^{\displaystyle
i\frac{\phi}{2}}(k^{(0)}_0+\frac{mc}{\hbar})\\ -e^{\displaystyle
i\frac{\phi}{2}}k^{(0)}_\_\\ -e^{\displaystyle
-i\frac{\phi}{2}}k^{(0)}_+
   \end{array}\right],
\end{equation}
where $\phi$ is the angle between $x^1$-axis and the particle's
polarization. By the similar method stated in the last example, we
get the solution in the particle's rest frame.
\begin{eqnarray}
\Psi|_{in\;rest\;frame} \cong \hspace*{9cm}\nonumber\\
\hspace*{-2cm}
\frac{\sqrt{2}/2}{\sqrt{\frac{2mc}{\hbar}(k^{(0)}_0+\frac{mc}{\hbar})}}
\left[\begin{array}{cccc} e^{\displaystyle
-i\frac{\phi}{2}}\{{(k^{(0)}_0+\frac{mc}{\hbar})}^2+{(k^{(0)}_+)}^2+\frac{ax^0}{c^2}k^{(0)}_+
(k^{(0)}_0+\frac{mc}{\hbar})\}\\ e^{\displaystyle
i\frac{\phi}{2}}\{{(k^{(0)}_0+\frac{mc}{\hbar})}^2+{(k^{(0)}_\_)}^2+\frac{ax^0}{c^2}k^{(0)}_\_
(k^{(0)}_0+\frac{mc}{\hbar})\}\\0\\0
   \end{array}\right]\nonumber\\
 \times e^{-i(\theta^{(0)}+\theta^{(1)}+\cdots)}|_{x^0=
 \frac{c^2}{a}\tanh^{-1}\frac{2v^{(0)}_{x^1}/c}{1+\frac{{v^{(0)}_{x^1}}^2}{c^2}}}.\hspace*{3.3cm}\nonumber
\end{eqnarray}
Supposing the particle's direction of spin is
$\hat{n}=(\cos\phi^\prime, \sin\phi^\prime,0)$, we solve the
equation
\begin{equation}
\hat{n}\cdot\hat{\Sigma}\;\;\Psi|_{in\;the\;rest\;frame}=\Psi|_{in\;the\;rest\;frame}
\end{equation}
where $\hat{\Sigma}^i=\left[\begin{array}{cc} \sigma^i \\0
\end{array} \begin{array}{cc} 0 \\ \sigma^i
\end{array}\right]$ is the spin operator of a Dirac particle. A
change in the polarization is found
\begin{equation}
\delta\phi=\phi^\prime-\phi\cong\frac{v^{(0)}_{x^1}v^{(0)}_{x^2}}{c^2}.
\end{equation}
This is different from the shift angle of a rod coming back to the
floor of the accelerated frame
\begin{eqnarray}
\delta\phi=\frac{v_{x^1}^{(0)}v_{x^2}^{(0)}}{c^2}(1-\cos2\phi).
\end{eqnarray}
 Obviously, its change depends on the angle
$\phi$. Why there is a difference between them?  When we studied
the trajectory of a Dirac particle, it occurred to us that a rod
can be viewed as composed of two Dirac particles. If there are two
particles placed in the accelerated frame, we can regard a
particle as one end and the other as the other end of the rod.
Since the trajectories of Dirac particles are the same as those of
scalar ones, the rod formed by two particles should have the same
rotation angle as a real rod's.  We think the point is that spin
is the internal structure of the particles but a rod is a
macroscopic object.  The vector of the spin polarization and a
vector in coordinate frame can't be regarded as the same thing.
Therefore, this should be one of the reasons for their difference.
In section 5 and Appendix B, a detailed discussion will be given.
\end{itemize}
\subsection{The trajectory of a Dirac particle in the accelerated frame}
So far, we have solved the Dirac equation approximately in the
uniformly accelerated frame and reached a consequence of a
rotation angle in polarization, when the particle falls back to
the ground. Subsequently, in this section, we will make an
investigation of the path of a Dirac particle in this frame and
see whether it follows the geodesic.\par Since at $x^0=0$ the
particle is at the origin of the frame, the initial wave vector
and the initial velocity
$\vec{v}(x^0=0)=(v^{(0)}_{x^1},v^{(0)}_{x^2},0)$ have the
relation:
\begin{eqnarray}
v^{(0)}_{x^1}/c &=& -k^{(0)}_1/k^{(0)}_0\nonumber\\
v^{(0)}_{x^2}/c &=& -k^{(0)}_2/k^{(0)}_0.
\end{eqnarray}
Suppose that ${\bf u}=\displaystyle \frac{d}{d \tau}$ is the four
velocity of the particle, then
\begin{equation}
k^\mu=\frac{m}{\hbar}\;u^\mu
\end{equation}
By using the chain rule:
\begin{equation}
\frac{dx^i}{dx^0}\;=\;\frac{\hbar/m}{(\displaystyle
\frac{dx^0}{d\tau})}\,k^i\;=\;\frac{k^i}{k^0}. \label{b2}
\end{equation}
Since we have known that $\displaystyle \frac{d x^i}{d x^0}$
approximates:
\begin{eqnarray}
\frac{d x^1}{d x^0} &\cong&
\frac{v^{(0)}_{x^1}}{c}-\frac{a}{c^2}x^0\label{b3}\\ \frac{d
x^2}{d x^0} &\cong& \frac{v^{(0)}_{x^2}}{c}\label{b4}\;.
\end{eqnarray}
Substitute (\ref{k}) into (\ref{b2}), differentiate them with
respect to $x^0$ and then substitute the major approximate terms
of ($d x^i/d x^0$), (\ref{b3}) and (\ref{b4}), into them, we can
obtain:
\begin{eqnarray}
\frac{d^2
x^1}{d{x^0}^2}&\cong&-\frac{a}{c^2}\{1+\frac{a}{c^2}(\frac{v^{(0)}_{x^1}}{c}x^0-\frac{1}{2}
\frac{a}{c^2}{x^0}^2)\}
+2\frac{a}{c^2}(\frac{v^{(0)}_{x^1}}{c}-\frac{a}{c^2}x^0)^2
\nonumber\\ \frac{d^2 x^2}{d{x^0}^2}&\cong&
2\frac{a}{c^2}\frac{v^{(0)}_{x^2}}{c}(\frac{v^{(0)}_{x^1}}{c}-\frac{a}{c^2}x^0).
\end{eqnarray}
These are the equation of motion which have been discussed in
section 1, so the trajectory of the a Dirac particle in the
accelerated frame obeys the same equation of a scalar particle to
$1/c^2$.\par
\subsection{A Dirac particle in the moving accelerated frame}
By the similar approximate method, eikonal approximation, now we
try to solve the non-inertial Dirac equation with the acceleration
$\vec{a}^\prime$ and the rotation $\vec{\omega}$. The equation:
\begin{eqnarray}
&&i\hbar\{\;\gamma^0\frac{\partial}{\partial
y^0}+(1+\varepsilon^2\frac{a^\prime
y^1}{c^2})\gamma^i\frac{\partial}{\partial y^i}
+\varepsilon^2\frac{a^\prime}{2c^2}\gamma^1-\varepsilon^3\frac{i}{c\hbar}\gamma^0\omega_3J_3\;\}\,\Psi\nonumber
\\&&=mc\,(1+\varepsilon^2\frac{a^\prime
y^1}{c^2})\,\Psi. \label{v}
\end{eqnarray}
where
\begin{equation}
J_3=\displaystyle y^1\frac{\hbar}{i}\frac{\partial}{\partial
y^2}-y^2\frac{\hbar}{i}\frac{\partial}{\partial
y^1}+\frac{\hbar}{2}\sigma^3.
\end{equation}
In the similar way, we assume
\begin{equation}
\Psi=(\Psi^{(0)}+\varepsilon\Psi^{(1)}+\varepsilon^2\Psi^{(2)}+\varepsilon^3\Psi^{(3)}+\cdots\;)\,e^{-i\theta}\label{u1}
\end{equation}
where $\Psi^{(0)}$ is a constant 4-component column vector,
$\Psi^{(1)}$ is a 4-component column vector function of the 1st
order $1/c$ of $y^o$,$y^1$ and $y^2$, $\Psi^{(2)}$ is a
4-component column vector function of the 2nd order $1/c$ of
$y^0$,$y^1$ and $y^2$, \ldots and so on.\par Here is a difference.
$\varepsilon$ represents the order $\frac{1}{c}$, not
$\frac{1}{c^2}$, since the additional term
$-\frac{i}{c\hbar}\gamma^0\omega_3J_3$ is of the order
$\frac{1}{c^3}$. As usual, $\theta$ is the phase of the Dirac
particle. In the inertial frame $\{\zeta^\mu\}$, the wave vector
is $\kappa_\mu^{(0)}$ and the phase
$\theta=\kappa^{(0)}_\mu\zeta^\mu$. After substituting the
coordinate transformation (\ref{gg}) into $\theta$,
\begin{equation}
\theta=\theta^{(0)}+\varepsilon\theta^{(1)}+\varepsilon^2\theta^{(2)}+\varepsilon^3\theta^{(3)}+\cdots
\;.
\end{equation}
where
\begin{eqnarray}
\theta^{(0)}&=&\kappa^{(0)}_0y^0+\kappa^{(0)}_1y^1+\kappa^{(0)}_2y^2\nonumber\\
\theta^{(1)}&=&0\nonumber\\
\theta^{(2)}&=&\frac{a^\prime}{2c^2}\kappa^{(0)}_1(y^0)^2+\frac{a^\prime}{c^2}\kappa^{(0)}_0y^0y^1\nonumber\\
\theta^{(3)}&=&-\frac{\omega_3}{c}\kappa^{(0)}_1y^2y^0+\frac{\omega_3}{c}\kappa^{(0)}_2y^1y^0.\nonumber\\
\vdots
\end{eqnarray}
Then the wave vector is
\begin{eqnarray}
\kappa_0&\cong&\kappa^{(0)}_0+\frac{a^\prime}{c^2}\kappa^{(0)}_1y^0+\frac{a^\prime}{c^2}\kappa^{(0)}_0y^1
-\frac{\omega_3}{c}\kappa^{(0)}_1y^2+\frac{\omega_3}{c}\kappa^{(0)}_2y^1\nonumber\\
& &\hspace*{.3cm}+\frac{{a^\prime}^2}{2c^4}\kappa^{(0)}_0{y^0}^2
+\frac{{a^\prime}^2}{c^4}\kappa^{(0)}_1y^1y^0-\frac{a^\prime\omega_3}{c^3}\kappa^{(0)}_0y^2y^0+\frac{{a^\prime}^3}{2c^6}\kappa^{(0)}_0y^1{y^0}^2\nonumber\\
\kappa_1&\cong&\kappa^{(0)}_1+\frac{a^\prime}{c^2}\kappa^{(0)}_0y^0+\frac{\omega_3}{c}\kappa^{(0)}_2y^0
+\frac{{a^\prime}^2}{2c^4}\kappa^{(0)}_1{y^0}^2+\frac{{a^\prime}^3}{6c^6}\kappa^{(0)}_0{y^0}^3\nonumber\\
\kappa_2&\cong&\kappa^{(0)}_2-\frac{\omega_3}{c}\kappa^{(0)}_1y^0-\frac{a^\prime\omega_3}{2c^3}\kappa^{(0)}_0{y^0}^2\nonumber\\
\kappa_3&=&0.\label{ad}
\end{eqnarray}
substituting (\ref{u}) into (\ref{v}), for the order of 1:
\begin{equation}
\{\;\gamma^0\theta^{(0)}_{,0}+\gamma^i\theta^{(0)}_{,i}-\frac{mc}{\hbar}\;\}\;\Psi^{(0)}=\;0
\end{equation}
and for the order $\varepsilon$:
\begin{equation}
i\gamma^0\Psi^{(1)}_{,0}+i\gamma^i\Psi^{(1)}_{,i}+\{\gamma^0\theta^{(0)}_{,0}+\gamma^i\theta^{(0)}_{,i}-\frac{mc}{\hbar}\;\}\;\Psi^{(1)}=\;0\label{y}
\end{equation}
Because we know that $\theta^{(1)}=0$ and the equation (\ref{v})
doesn't have any term of the order $\varepsilon$, $\Psi^{(1)}$
should not have any contribution for $\Psi$. So
\begin{equation}
\Psi^{(1)}=0  .
\end{equation}
The equation of the next order $\varepsilon^2$ is
\begin{eqnarray}
i\gamma^0\Psi^{(2)}_{,0}+i\gamma^i\Psi^{(2)}_{,i}+(\gamma^0\theta^{(0)}_{,0}+
\gamma^i\theta^{(0)}_{,i}-\frac{mc}{\hbar})\Psi^{(2)}\hspace{3.5cm}\nonumber\\
=-\;(\gamma^0\theta^{(2)}_{,0}+\gamma^i\theta^{(2)}_{,i})\;\Psi^{(0)}
-\frac{a^\prime
y^1}{c^2}(\gamma^i\theta^{(0)}_{,i}-\frac{mc}{\hbar})\Psi^{(0)}-i\frac{a^\prime}{c^2}\gamma^1
\label{z} \Psi^{(0)}.\label{w}
\end{eqnarray}
This equation has the same form as the equation (\ref{j}),
naturally they have the same solutions in the same initial
conditions. So the problem is to solve the equation to the next
order. The equation of the order $\varepsilon^3$ with substituting
the partial derivatives of  $\theta^{(0)}$ and $\theta^{(3)}$ into
the above equation becomes
\begin{eqnarray}
i\gamma^0\Psi^{(3)}_{,0}+i\gamma^i\Psi^{(3)}_{,i}+(\gamma^0\theta^{(0)}_{,0}+
\gamma^i\theta^{(0)}_{,i}-\frac{mc}{\hbar})\Psi^{(3)}\hspace*{3cm}\nonumber\\\
=-\frac{\omega_3}{c}y^0(\gamma^1\kappa^{(0)}_2-\gamma^2\kappa^{(0)}_1)\Psi^{(0)}-\frac{\omega_3}{2c}\gamma^0\sigma^3\Psi^{(0)}.\label{x}
\end{eqnarray}
Here $\Psi^{(2)}$ can be solved yourself and will be omitted
without losing completeness.
\subsection{The trajectory of a Dirac particle in the moving accelerated frame}
In the section 4.2, we have proved that, in a uniformly
accelerated frame, a Dirac particle will follow the geodesic.
Moreover, according to SREP, the trajectory of a flying clock in
the moving accelerated frame is slightly deviated from that in the
uniformly accelerated frame. In this section, we will follow the
procedures in the section 4.2 and demonstrate that the trajectory
of a Dirac particle in the moving accelerated frame also obeys the
modified equation of motion added the terms of gravitomagnetic
accelerations.\par By the same way as given in the section 4.2, we
can obtain finally:
\begin{eqnarray}
\frac{d^2
y^1}{d{y^0}^2}&\cong&-\frac{a^\prime}{c^2}(1+\frac{a^\prime}{c^2}(\frac{v^{(0)}_{y^1}}{c}y^0-\frac{1}{2}
\frac{a^\prime}{c^2}{y^0}^2))+2\frac{\omega_3}{c}\frac{v^{(0)}_{y^2}}{c}
+2\frac{a^\prime}{c^2}(\frac{v^{(0)}_{y^1}}{c}-\frac{a^\prime}{c^2}y^0)^2
\nonumber\\ \frac{d^2
y^2}{d{y^0}^2}&\cong&-2\frac{\omega_3}{c}(\frac{v^{(0)}_{y^1}}{c}-\frac{a^\prime}{c^2}y^0)
+2\frac{a^\prime}{c^2}\frac{v^{(0)}_{y^2}}{c}(\frac{v^{(0)}_{y^1}}{c}-\frac{a^\prime}{c^2}y^0).
\end{eqnarray}
Like the results in the section 4.2, the Dirac particle in the
moving accelerated frame also complies with the modified equation
of motion, the geodesic equation. Hence, we know that the
trajectory of a Dirac particle in the accelerated frame, or in the
moving accelerated frame, is the same as that of a scalar particle
to $1/c^2$, when they are given the same initial conditions.
\section{Discussion}
The trajectories of Dirac particles in the accelerated frame and
the moving accelerated frame are found not to be different from
those of scalar particles. However, this result is desirable from
equivalence principle. Equivalence principle tells us that in a
gravitational field the trajectory of a test body with given
initial conditions is independent of its internal structure and
composition,. If equivalence principle is correct, the
trajectories of all kinds of particles with the same initial
conditions shouldn't differ from each other.\par The rotation
angle of a rod thrown in an accelerated frame has been discovered
by K. Nordtvedt \cite{n}. The rotation angle of the rod is
\begin{eqnarray}
\delta \phi = \frac{v^{(0)}_{x^1}v^{(0)}_{x^2}}{c^2}(1-\cos
2\phi).
\end{eqnarray}
In section 4, we obtained the similar angle of a Dirac particle
falls back to the floor. It is
\begin{eqnarray}
\delta \phi = \frac{v^{(0)}_{x^1}v^{(0)}_{x^2}}{c^2}.
\end{eqnarray}
Since we know that the polarization of spin is an internal
structure of Dirac particles, it should not be affected by the
factors which can only change the microscopic phenomena. In
appendix B, the part of angular dependence of the rotation angle
can be explained by Lorentz contraction. Hence, the rotation
angles of a rod and a Dirac particle are consistent with each
other.\par The rotation of spin due to moving acceleration is the
same as the rotation of angular momentum for a Dirac particle.
Hence the gyro-gravitational factor is one for moving
acceleration.

\appendix
\section{}
\noindent
 In the
section 2, we showed the transformation from $\{y^\mu\}$ to
$\{\zeta^\mu\}$ is (\ref{gg}). However, how do we make sure if
these terms are correct? Since we used the Lorentz transformation
to approximate the coordinate transformation from $\{y^\mu\}$ to
$\{x^\mu\}$. In this appendix, we will demonstrate the correctness
of this transformation to the order $\frac{1}{c^2}$. Then we can
believe the terms in (\ref{gg}) we chosen are correct.\par In
fact,the terms in (\ref{gg}) are pick from the successive
transformations from the moving accelerated frame, via the
accelerated frame, to an inertial frame. With setting $y^0=ct_y$
and $\zeta^0=ct_\zeta$, the transformation becomes
\begin{eqnarray}
t_\zeta&=&(1+\frac{w^4}{c^4}+\cdots)t_y+(1+\frac{w^2}{c^2}+\cdots)\frac{a}{c^2}y^1t_y+(1+\frac{w^2}{c^2}+\cdots)\frac{aw}{c^4}y^1y^2\hspace*{3cm}
\nonumber\\
&&+\frac{a^2}{6c^2}(1+2\frac{w^2}{c^2}+\cdots){y^0}^3+\frac{a^2}{2c^4}(1+2\frac{w^2}{c^2}+\cdots)w{t_y}^2y^2\nonumber\\
&&+\frac{a^2}{2c^4}(1+2\frac{w^2}{c^2}+\cdots)\frac{w^2}{c^2}t_y{y^2}^2+\frac{a^2}{6c^4}(1+2\frac{w^2}{c^2}+\cdots)\nonumber\\
&&\frac{w^3}{c^4}{y^2}^3+\frac{a^3}{6c^4}(1+2\frac{w^2}{c^2}+\cdots){t_y}^3y^1+\frac{a^3}{2c^6}(1+2\frac{w^2}{c^2}+\cdots)wy^1y^2{t_y}^2\nonumber\\
&&+\frac{a^3}{2c^6}(1+2\frac{w^2}{c^2}+\cdots)\frac{w^2}{c^2}t_yy^1{y^2}^2+\frac{a^3}{6c^6}(1+2\frac{w^2}{c^2}+\cdots)\frac{w^3}{c^4}y^1{y^2}^3+\cdots\nonumber\\
\zeta^1&=&y^1+\frac{a}{2}(1+\frac{w^2}{c^2}+\cdots){t_y}^2+\frac{aw}{c^2}(1+\frac{w^2}{c^2}+\cdots)t_yy^2\nonumber\\
&&+\frac{aw^2}{2c^4}(1+\frac{w^2}{c^2}+\cdots){y^2}^2+\frac{a^2}{2c^2}(1+\frac{w^2}{c^2}+\cdots)y^1{t_y}^2\nonumber\\
&&+\frac{a^2w}{c^3}(1+\frac{w^2}{c^2}+\cdots)t_yy^1y^2+\frac{a^2w^2}{2c^6}(1+\frac{w^2}{c^2}+\cdots)y^1{y^2}^2+\cdots\nonumber\\
\zeta^2&=&(1+\frac{w^4}{c^4}+\cdots)y^2-\frac{aw}{c^2}(1+\frac{w^2}{c^2}+\cdots)y^1t_y\nonumber\\
&&-\frac{aw^2}{c^4}(1+\frac{w^2}{c^2}+\cdots)y^1y^2+\cdots\hspace*{1.8cm}\nonumber\\
\zeta^3&=&y^3.\label{hh}
\end{eqnarray}
Therefore, our purpose is to judge which term is right. Since only
the transformation from the moving accelerated frame to uniformly
accelerated frame is approximate, we need to check the validity of
the Lorentz transformation which is regarded as the transformation
from $\{y^\mu\}$ to $\{x^\mu\}$. The Lorentz transformation to the
order $\frac{1}{c^2}$ is (\ref{u}). Now we should prove that no
other correction term of the order $\frac{1}{c}$ and
$\frac{1}{c^2}$ exists.\par Because there are only three
parameters, $a$, $w$ and $c$, appear in our discussion, the
coefficients of $t_y$ and $y^i$ must be the combination of these
three parameters. When $w=0$, this transformation is an identity
transformation so that the possible additional terms must be in
the form $\frac{a^nw^m}{c^p}$, where $n$, $m$, $p$ are integers
and $n\neq0$, $m\neq0$. By considering the dimension of the
possible terms, only four terms may exist:
\begin{eqnarray}
t_x&\cong&(1+\frac{w^2}{2c^2})t_y-\frac{w}{c}y^2+A\frac{aw}{c}{t_y}^2+B\frac{aw}{c^2}{t_y}^3\nonumber\\
x^1&=&y^1\nonumber\\
x^2&\cong&(1+\frac{w^2}{2c^2})y^2-w(1+\frac{w^2}{2c^2})t_y+C\frac{aw}{c^2}y^1t_y+D\frac{aw}{c^2}y^2t_y\nonumber\\
x^3&=&y^3,
\end{eqnarray}
where A, B, C and D are dimensionless constant. If we are the
observers in the frame $\{x^\mu\}$, we will observe the frame
$\{y^\mu\}$ leaving us with the constant velocity $w$ in
$x^2$-direction. Suppose the object which is at rest in
$\{y^\mu\}$ stays at the position $(y^1_c,y^2_c,y^3_c)$. The path
of the object can be described by the parameter $t_y$:
\begin{eqnarray}
t_x(t_y)&\cong&(1+\frac{w^2}{2c^2})t_y+A\frac{aw}{c}{t_y}^2+B\frac{aw}{c^2}{t_y}^3-\frac{w}{c}y^2_c\nonumber\\
x^1(t_y)&=&y^1_c\nonumber\\
x^2(t_y)&\cong&\{-w(1+\frac{w^2}{2c^2})+\frac{aw}{c^2}(Cy^1_c+Dy^2_c)\}t_y+(1+\frac{w^2}{2c^2})y^2_c\nonumber\\
x^3(t_y)&=&y^3_c.
\end{eqnarray}
It is obvious that $A$=$B$=$0$, because the terms of the constants
$A$ and $B$ are not linear. The terms of $C$ and $D$ are linear,
but they depend on the position of the object. Therefore,
$C$=$D$=$0$. Then we have verified the approximate transformation
(\ref{gg}), Lorentz transformation, is correct to the order
$\frac{1}{c^2}$.\par Now we have reasons to believe the terms of
the order $\frac{1}{c^2}$ in (\ref{hh}) are correct, and they are
just the ones in (\ref{gg}) and are in agreement with the
consequence of Li \& Ni \cite{b}.
\section{}
\noindent From Nordtvedt's paper \cite{n}, he discovered that in a
uniformly accelerated frame a rod with the initial velocity
$\vec{v}(x^0=0)$=$(v^{(0)}_{x^1},v^{(0)}_{x^2},0)$ will rotate an
angle $\delta\phi$, when it falls back to the floor. He got
\begin{equation}
\delta\phi=\frac{v^{(0)}_{x^1}v^{(0)}_{x^2}}{c^2}(1-\cos2\phi).\label{B1}
\end{equation}
The first term can be derived from Thomas precession
\begin{equation}
\frac{d\phi}{dt}=\frac{1}{2}\frac{gv^{(0)}_{x^2}}{c^2}.
\end{equation}
In Appendix B, we will illuminate that second term comes from
Lorentz contraction. When an object moves, according to special
relativity, it's length will contract. Naturally, the rod in our
discussion will, too. Suppose in the inertial frame $\{t_z,z^i\}$
the rod is at rest. If there exists another frame $\{t_x,x^i\}$,
the rod leaves the frame with the velocity $\vec{v}$=
$(v_1,v_2,0)$. The trajectory of the rod's two ends are
\begin{eqnarray}
x^1_A(t_x)=v_1t_x &\textup{and}&
x^1_B(t_x)=v_1t_x+L^\prime_1\nonumber\\ x^2_A(t_x)=v_2t_x
&\textup{and}& x^2_B(t_x)=v_2t_x+L^\prime_2,
\end{eqnarray}
where $L_1^\prime$ and $L_2^\prime$ are constants, and in its rest
frame $\{t_z,z^i\}$ are
\begin{eqnarray}
z^1_A(t_z)=0 &\textup{and}& z^1_B(t_z)=L_1\nonumber\\ z^2_A(t_z)=0
&\textup{and}& z^2_B(t_z)=L_2,\label{aaa}
\end{eqnarray}
where $L_1$ and $L_2$ are also constants. The coordinate
transformation from $\{t_z,z^i\}$ to $\{t_x,x^i\}$ is
\begin{eqnarray}
t_x&=&\gamma(t_z+\frac{v_1}{c^2}z^1+\frac{v_2}{c^2}z^2)\nonumber\\
x^1&=&\{1+(\gamma-1)\frac{\frac{{v_1}^2}{c^2}}{\frac{{v_1}^2}{c^2}+\frac{{v_2}^2}{c^2}}\}z^1+(\gamma-1)\frac{\frac{v_1v_2}{c^2}}{\frac{{v_1}^2}{c^2}+\frac{{v_2}^2}{c^2}}z^2+\gamma
v_1t_z\nonumber\\
x^2&=&\{1+(\gamma-1)\frac{\frac{{v_2}^2}{c^2}}{\frac{{v_1}^2}{c^2}+\frac{{v_2}^2}{c^2}}\}z^2+(\gamma-1)\frac{\frac{v_1v_2}{c^2}}{\frac{{v_1}^2}{c^2}+\frac{{v_2}^2}{c^2}}z^1+\gamma
v_2t_z\nonumber\\ x^3&=&z^3,\label{ccc}
\end{eqnarray}
where $\gamma$=$\displaystyle
\frac{1}{\sqrt{1-\frac{{v_1}^2}{c^2}-\frac{{v_2}^2}{c^2}}}$. At
$t_x$=$T_x$, the end $A$ in the rest frame is at the time $T_z^A$,
and
\begin{equation}
T_z^A=\gamma^{-1}T_x.
\end{equation}
As to the end $B$, it is at the time $T_z^B$, and
\begin{equation}
T^B_z=\gamma^{-1}T_x-\frac{v_1}{c^2}L_1-\frac{v_2}{c^2}L_2.
\end{equation}
Hence,
\begin{equation}
T^B_z=T_z^A-\frac{v_1}{c^2}L_1-\frac{v_2}{c^2}L_2.\label{bbb}
\end{equation}
Therefore, at $t_x$=$T_x$, the end $A$ in the frame $\{t_x,x^i\}$
is at the position:
\begin{eqnarray}
x^1_A(T_x)&=&\gamma v_1T^A_z\nonumber\\ x^2_A(T_x)&=&\gamma
v_2T^A_z.\label{eee}
\end{eqnarray}
Substituting (\ref{aaa}) and (\ref{bbb}) into the coordinate
transformation (\ref{ccc}), we can obtain the end $B$'s position
at $t_x$=$T_x$:
\begin{eqnarray}
x^1_B(T_x)&=&\{1+(\gamma-1)\frac{\frac{{v_1}^2}{c^2}}{\frac{{v_1}^2}{c^2}+\frac{{v_2}^2}{c^2}}\}L_1+(\gamma-1)\frac{\frac{v_1v_2}{c^2}}{\frac{{v_1}^2}{c^2}+\frac{{v_2}^2}{c^2}}L_2\nonumber\\
&&+\gamma
v_1(T_z^A-\frac{v_1}{c^2}L_1-\frac{v_2}{c^2}L_2)\nonumber\\
x^2_B(T_x)&=&\{1+(\gamma-1)\frac{\frac{{v_2}^2}{c^2}}{\frac{{v_1}^2}{c^2}
+\frac{{v_2}^2}{c^2}}\}L_2+(\gamma-1)\frac{\frac{v_1v_2}{c^2}}{\frac{{v_1}^2}{c^2}
+\frac{{v_2}^2}{c^2}}L_1\nonumber\\
&&+\gamma
v_2(T_z^A-\frac{v_1}{c^2}L_1-\frac{v_2}{c^2}L_2).\nonumber\\
\label{ddd}
\end{eqnarray}
Now the orientation of the rod in the frames $\{t_x,x^i\}$ and
$\{t_z,z^i\}$ can be compared. In the rod's rest frame, we define
\begin{equation}
\tan\phi\equiv\frac{z^1_B-z^1_A}{z^2_B-z^2_A}=\frac{L_1}{L_2}.
\end{equation}
In the similar way, the tangent of the angle in the frame
$\{t_x,x^i\}$ is
\begin{equation}
\tan\phi^\prime=\frac{x^1_B-x^1_A}{x^2_B-x^2_A}=\frac{L^\prime_1}{L^\prime_2}.\label{fff}\\
\end{equation}
By substituting (\ref{eee}) and (\ref{ddd}) into (\ref{fff}), the
tangent of $\phi^\prime$ is obtained:
\begin{equation}
\tan\phi^\prime\cong\frac{L_1}{L_
2}\{1-(\frac{{v_1}^2}{c^2}-\frac{{v_2}^2}{c^2})
-\frac{v_1v_2}{c^2}(\frac{{L_2}^2-{L_1}^2}{L_1L_2})\}.
\end{equation}
Finally, the rotation angle due to Lorentz contraction is
\begin{equation}
\delta\phi''=\phi^\prime-\phi\cong-\frac{1}{2}(\frac{{v_1}^2}{c^2}-\frac{{v_2}^2}{c^2})\frac{L_1L_2}{{L_1}^2+{L_2}^2}
-\frac{1}{2}\frac{v_1v_2}{c^2}\frac{{L_1}^2-{L_2}^2}{{L_1}^2+{L_2}^2}.
\end{equation}
We can go back to the problem of the rod's the rotation angle in
the accelerated frame. Since the velocity of the rod changes from
$(v^{(0)}_{x^1},v^{(0)}_{x^2},0)$ to
$(-v^{(0)}_{x^1},v^{(0)}_{x^2},0)$, the rod will rotate:
\begin{eqnarray}
\delta\phi_{\{contraction\}}&=&\delta\phi''(v_1=-v^{(0)}_{x^1},v_2=v^{(0)}_{x^2})-\delta\phi''(v_1=v^{(0)}_{x^1},v_2=v^{(0)}_{x^2})\nonumber\\
&\cong&\;\;\;\frac{v^{(0)}_{x^1}v^{(0)}_{x^2}}{c^2}\frac{{L_1}^2-{L_2}^2}{{L_1}^2+{L_2}^2}\nonumber\\
&=&\;\;\;-\frac{v^{(0)}_{x^1}v^{(0)}_{x^2}}{c^2}\cos2\phi.
\end{eqnarray}
This is the second term of (\ref{B1}).

\end{document}